\documentclass[review]{elsarticle}

\usepackage{lineno,hyperref}
\usepackage{amsmath}
\usepackage{amssymb}
\usepackage{mathdots}
\usepackage{graphicx}
\usepackage{siunitx}
\usepackage{algorithmic}
\usepackage{textcomp}
\usepackage{caption}
\usepackage{subcaption}
\usepackage{cuted}

\DeclareSIUnit{\count}{count}

\newtheorem{Remark}{Remark}

\newtheorem{Theorem}{Theorem}

\journal{arXiv}
	
\begin{document}

\begin{frontmatter}

\title{Extension of a Linear Controller Scheme to Non-Linear Systems and its Application on Inverted Pendulum}

\author{Justin Jacob\corref{mycorrespondingauthor}}
\cortext[mycorrespondingauthor]{Corresponding author}
\ead{justinjacob@iitb.ac.in}
\author{Navin Khaneja}
\ead{nkhaneja@iitb.ac.in}
\address{Systems and Control Engineering Department, \\Indian Institute of Technology, Bombay}

\begin{abstract}
This paper presents the control and stabilization of the rotary inverted pendulum based on a general controller scheme. The proposed scheme has its foundation in classical control theory, and the importance of an integrator in disturbance rejection is emphasized. The system's dynamics are obtained by the Euler Lagrange method and are approximated for small-angle as balancing the pendulum is the objective. Experimental results demonstrate that the proposed control scheme can achieve the stabilization of a non-linear system. Also, the boundedness and convergence of the non-linear system with the controller subjected to the initial condition are validated.
\end{abstract}


\end{frontmatter}


\section{Introduction}

The development of new control strategies and control theories evolve from a few fundamental problems. For researchers in robotics and mechatronics, the inverted pendulum problem is a fundamental benchmark problem. The inverted pendulum has a simple structure regardless of its highly non-linear dynamics, encouraging the researchers to apply various control schemes and analyze them. A wide range of control strategies in the literature for the control of inverted pendulum can be found. Some popular techniques used are bang-bang control\cite{239008}, Fuzzy logic control\cite{481841}, PID Adaptive control\cite{CHANG2002233}, Sliding mode control\cite{sliding_pendulum}, Time optimal control\cite{Chernousko2007}.

The paper presents the control and stabilization of the non-linear rotary inverted pendulum. The stabilization of the pendulum arm is achieved with the help of the rotary arm, which is manipulated by an actuator. The system has two equilibrium points, considering the rotating arm to be stationary. The unstable equilibrium point corresponds to the upright position of the pendulum, and stabilizing the pendulum arm at this point with the help of a general control scheme is presented in this paper. The proposed controller is model-based, and the system's dynamics are obtained from the Euler Lagrange equation.

The essence of the proposed controller lies in the classical control theory. The controller can eliminate any bounded disturbance and stabilize the linear system. The importance of the integral controller in disturbance rejection is emphasized in this paper. The significance of modelling disturbance in model-based controllers is well described in the literature\cite{1223451}. In this paper, we model the disturbance as a sequence of step inputs, which is the crucial idea in the disturbance rejection controller.  When the controller is applied to the non-linear system, the non-linearities are postulated as the disturbance to the system. The controller parameters are designed from the linear part of the system. Stabilization of the non-linear system can be achieved with the proposed controller, subjected to initial conditions. The boundedness and convergence of the system with the controller are shown. The paper presents an effective way of attaining the controller parameters.

The proposed controller is implemented on the rotary inverted pendulum in a hardware-in-the-loop fashion. The results corresponding to these demonstrate that the controller can stabilize and control the non-linear system. 

\section{Theory}

\subsection{Disturbance Formulation}
\begin{figure}[!h]
\begin{center}
\includegraphics[scale=.3]{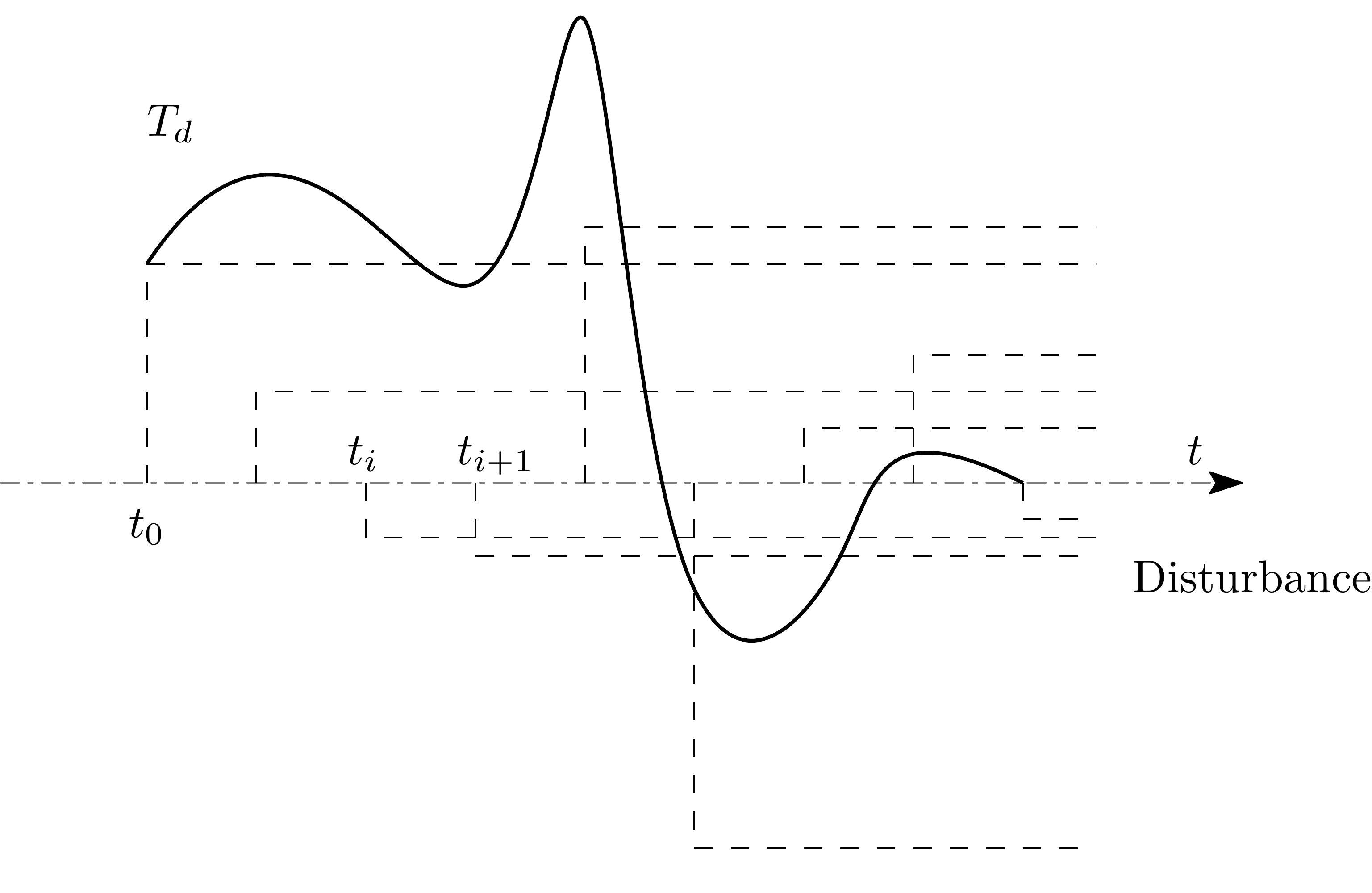}
\caption{Approximation of the disturbance with step functions.}
\label{disturbance}
\end{center}
\end{figure}

One of the most important parts of the controller design is to model a disturbance and design the controller to eliminate these. Figure \ref{disturbance} shows a random disturbance, $T_d$ with $\| T_d \| < \infty $. 

Any function of the form shown in figure \ref{disturbance} can be approximated by a sequence of step functions as

\begin{equation}
	T_d \sim \sum_{i} \alpha_i r(t-t_i)
\end{equation}
where $r(t) = 1$ for $t \geq 0$ and $\alpha_i \in \mathbb{R}$ are arbitrary constants. Theoretically

\begin{equation}
	T_d(t) = \lim_{(t_{i+1}-t_i) \to 0} \sum_{i} \alpha_i r(t-t_i).
\end{equation} 
We first propose a controller that can eliminate this sequence of step disturbance acting on a linear system, then extend it to the application of a non-linear system. Since we are considering step disturbance, it's similar to showing the system stabilizes for a step input.

\begin{Remark}\label{remark1}
\textbf{Theoretically, every disturbance can be modelled by a sequence of step input, and an integrator in the controller eliminates it.}\\
\end{Remark}

\subsection{Controller Design}

An $n^{th}$ order general linear time-invariant (LTI) system can be described by,

\begin{equation}\label{generalsystem}
\dfrac{d^n x(t)}{dt^n} + \sum_{i=1}^{n} a_i \dfrac{d^{(i-1)} x(t)}{dt^i} = u(t) + T_d.
\end{equation}
The model is taken as it resembles the inverted pendulum dynamics. Here $u(t) \in \mathbb{R}$ is the input, and we assume a single output which is the state, $x(t) \in \mathbb{R}$, and other states as the successive derivatives. $T_d \in \mathbb{R}$ is the disturbance acting on the system, with $\| T_d \| < \infty $.

\begin{Theorem} \label{objective}

Any system of the form Eq.\eqref{generalsystem} can be stabilized using the control input,  \\
\begin{equation}\label{controller}
u(t) = b_0 \int_0^t z(t) \, dt + \sum_{i=1}^{n} b_i \dfrac{d^{(i-1)} z(t)}{dt^{i-1}}.
\end{equation}
subjected to 
$a_i + b_i , \, b_0 > 0$, $\| u \| \leq u_{max}$ and $b_i$ ensuring the system characteristics equation to be Hurwitz\cite{chen1999linear}. Where $z(t) = x_d - x(t)$ with $x_d \in \mathbb{R}$ as the desired output and $b_i \in \mathbb{R}$ as the gain constants corresponding to the states. 

\end{Theorem}

\begin{Remark}\label{remark2}
\textbf{For a system of $n^{th}$ order, $(n~-~1)$ derivatives, a proportional and an integral controller part are necessary for the control law to control, stabilize, and reject disturbance. }
\end{Remark}

\subsection{Proof}

\subsubsection{Disturbance Rejection}

\begin{figure}[!h]
\centering
\includegraphics[scale=.3]{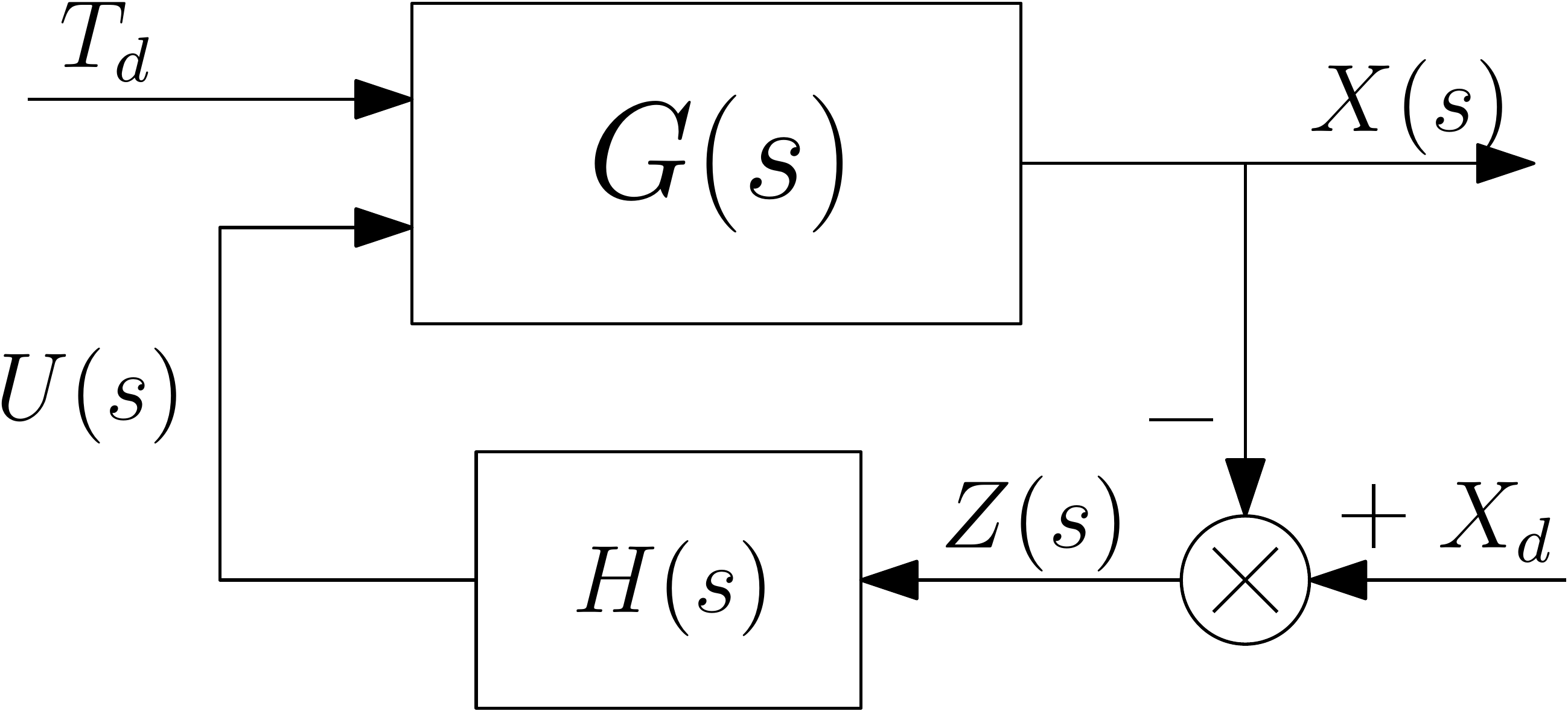}
\caption{Model of the effective system.}
\label{model}
\end{figure} 
The output of the effective system is obtained as

\begin{equation} \label{disturbance_op}
 X(s) = \frac{\sum_{i} \alpha_i e^{-s t_i}}{s} \left(\frac{G(s)}{1 + G(s) H(s)} \right) + \frac{x_d}{s} \left(\frac{G(s) H(s)}{1 + G(s) H(s)} \right).
\end{equation}
$H(s)$ is chosen to stabilize the system from the disturbance, which is given by the first part in Eq.\eqref{disturbance_op}. As the system is linear, it is sufficient to show the controller eliminates the disturbance corresponding to a step input. The main intention is to eliminate the existence of $\frac{1}{s}$ from the output equation. With $H(s) = \frac{b_0}{s} $, which in turn tells $u = b_0 \int_0^t z(\tau) d\tau$, an \textbf{integral controller}, eliminates $\frac{1}{s}$ term. The key idea is to represent the disturbance as a sequence of step inputs. For small-time $t = \epsilon$ one can obtain the sequence step functions which will resemble the disturbance. And these step responses will decay as the system considered is a stable one. The response to one of the step input is

\begin{equation}\label{finalvalue}
X(s) = \dfrac{\alpha_0 + b_0 \, x_d/s}{s^{n+1} + a_n s^n + \cdots + a_1 s + b_0}.
\end{equation}  
Applying the final value theorem\cite{fvt_ivt} on Eq.\eqref{finalvalue}, $x$ $\longrightarrow x_d$ as $t$~$\longrightarrow~\infty$, as long as $b_0$ make the system Hurwitz. Most of the practical disturbance have a span very much less than the operational period. Hence all it's effect, will be eliminated over time.  

\begin{Remark}\label{remark3}
\textbf{The system with the new integral state, must also be controllable.}\\
\end{Remark}

\subsubsection{Stabilization}

A stable system is not always guaranteed; hence the first assumption of stability does not hold in every case. The system is made stable by the idea of pole placement by feedback\cite{jacobpole}. The key idea here is to manipulate each coefficient in the denominator of the system. 

\begin{equation}\label{stablecontroller}
X(s) = \dfrac{\dfrac{x_d}{s}\left(\dfrac{1}{s^n + \sum_{i=1}^{n} a_is^{i-1}} \right)H(s)}{1+ \left(\dfrac{1}{s^n + \sum_{i=1}^{n} a_is^{i-1}} \right) H(s)}
\end{equation}
$H(s)$ has to account for all the coefficients corresponding to $s^0$ to $s^{n-1}$. Thus, $H(s) = b_1 s^0 + b_2 s^1 + \cdots + b_n s^{n-1} $,  and the input $ u = b_1 y + b_2 \dfrac{dy}{dt} + b_3 \frac{d^2 y}{dt^2} + \cdots + b_{n} \dfrac{d^{n-1} y}{dt^{n-1}} $, which is a combination of \textbf{derivative controllers}. Taking proportional term as the zeroth derivative, this shows \textbf{a one-to-one relationship between the number of derivatives to the order of the system}. Combining the disturbance rejection controller and the stabilization controller the general controller scheme is obtained. Applying the general controller scheme to Eq.\eqref{stablecontroller}, the effective denominator becomes

\begin{equation}
s^{n+1} + \sum_{i=1}^{n} (a_i+b_i)s^{i} + b_0 .
\end{equation}
By pole placement the system can be made Hurwitz hence giving us a stable system, and corresponding gain values. When taking the Laplace transform with non-zero initial conditions, an additional term in the numerator appears corresponding to the initial values. This term makes a proper fraction where the denominator power is greater than the numerator, and hence its effect goes to zero as $t$~$\longrightarrow~\infty$.

\section{Pendulum Model}

\begin{figure}[!h]
\begin{center}
\includegraphics[scale=.25]{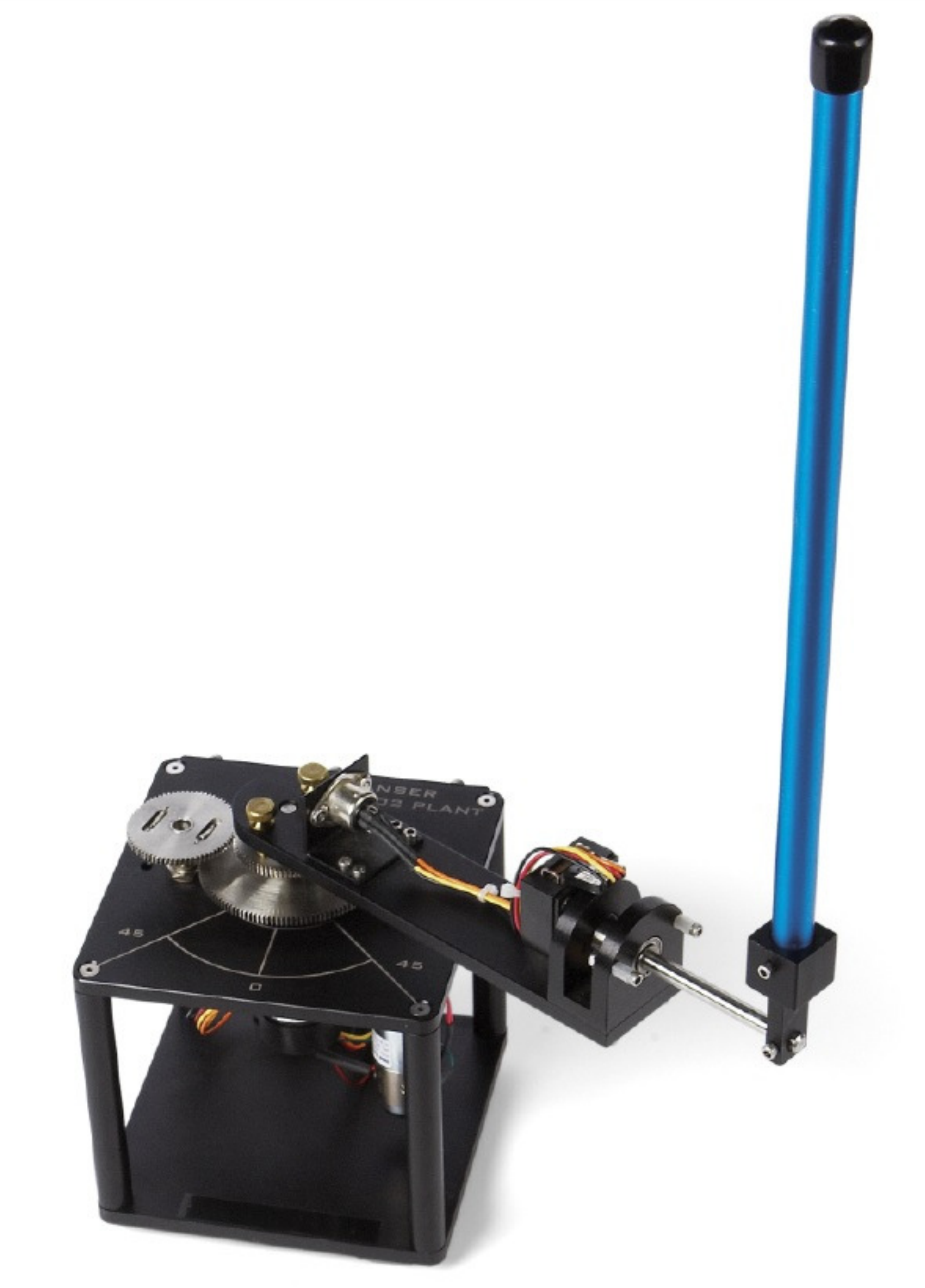}
\caption{Rotary inverted pendulum set-up}
\end{center}
\label{rip_setup}
\end{figure}

The rotary inverted pendulum mainly consists of a rotary arm actuated by a servo motor and a pendulum connected to the rotary arm. The rotary arm has a length of $L_1$, the moment of inertia about centre of mass of $J_1$, and a mass of $M_1$. The angle between the rotary arm and the $X$-axis (horizontal plane) is taken as $\theta$. The pendulum has a length of $L_2$, the moment of inertia about centre of mass of $J_2$, and a mass of $M_2$. The angle between the pendulum and the $Z_0$-axis ($Z_0 Y_0$ plane) is taken as $\alpha$. By convention counter clock wise (CCW) motion is taken as a positive angle. Center of mass of the pendulum's mass occurs at $L_2/2$ and that of the rotary arm at $\SI{.0619} {\meter}$ of the total length $L_1 = \SI{.216} {\meter}$. The ratio $\dfrac{.0619}{.216} = .2865$ is approximated to $\dfrac{2}{7} = .2857$.

\begin{figure}[!h]
\begin{center}
\includegraphics[scale=.7]{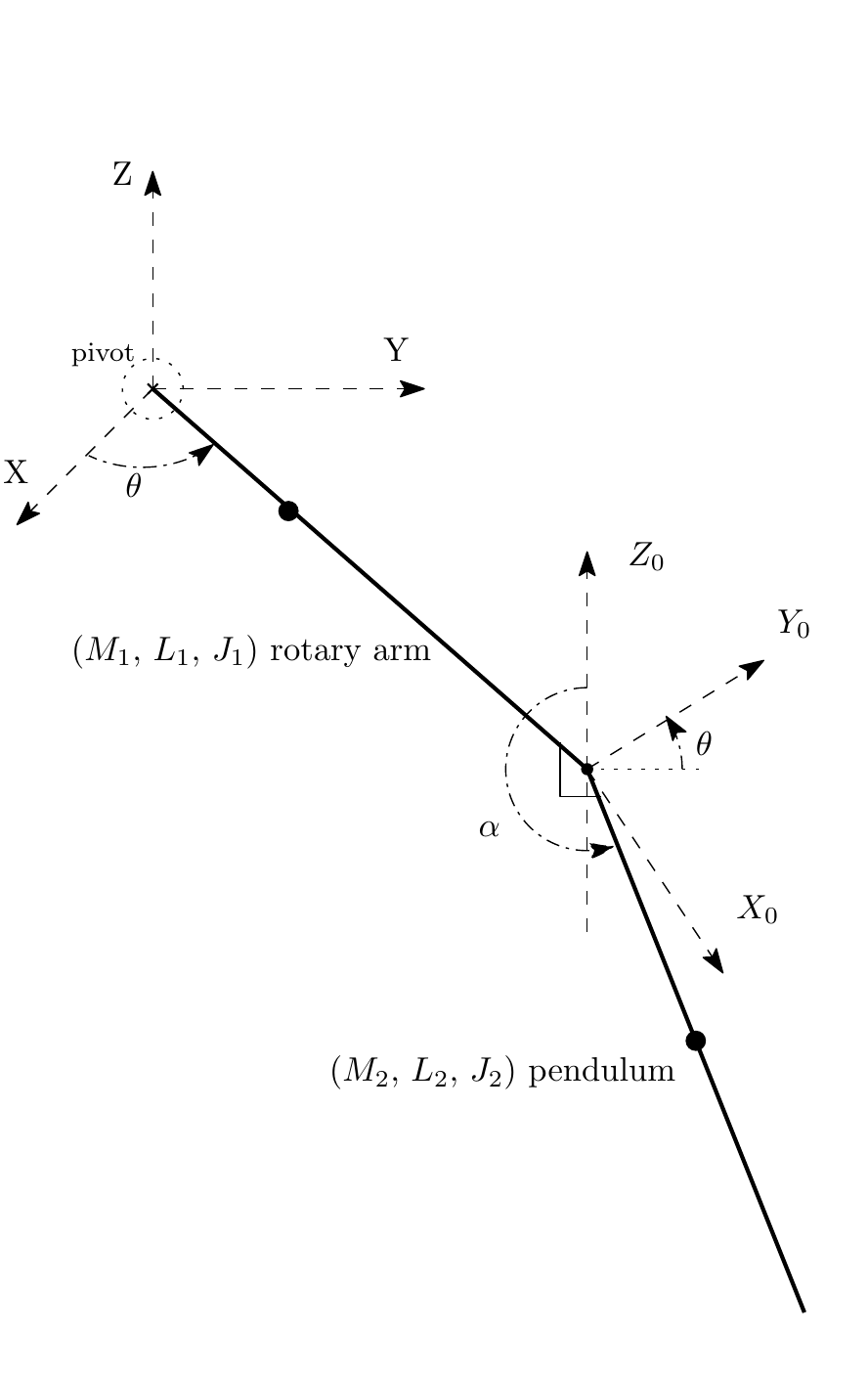}
\caption{Rotary inverted pendulum model}
\end{center}
\label{rip_model}
\end{figure}

The Euler-Lagrange's method\cite{morin2008introduction} is used to obtain the dynamic model of the system where the Lagrangian coordinates ($q$) are $\theta$ and $\alpha$ and torque ($F$) acting on rotary arm as the non conservative force. From Euler Lagrange's equation $\dfrac{d}{dt}\dfrac{\partial L}{\partial \dot{q_i}} - \dfrac{\partial L}{\partial q_i} + \dfrac{\partial D}{\partial \dot{q_i}} = F_i$ where $L$ is the lagrangian of the system (Total energy = potential energy (P.E) + kinetic energy (K.E)) and $D$ is the Rayleigh dissipation function (viscous friction forces). $D = \dfrac{1}{2} B_1 \dot{\theta^2} + \dfrac{1}{2} B_2 \dot{\alpha^2}$ where $B_1$ and $B_2$ are the yaw and pitch viscous friction thrust coefficients respectively. 

\section{Equation of Motion}

Potential energy due to rotary arm, let the height of the rotary arm from the ground be $h$, then potential energy is $M_1 g h$. The change in the potential energy of the rotary arm due to a change in $\theta$ is $0$. Potential energy due to pendulum is $M_2 g \left( h + \dfrac{L_2}{2} \cos\alpha \right)$. The changing part of potential energy of pendulum due to change in $\alpha$ is $M_2 g \dfrac{L_2}{2} \cos\alpha$. Here the reference of $\alpha$ is taken with respect to the upright position of the pendulum.  \\

Total potential energy 

\begin{equation}
PE = M_2 g \dfrac{L_2}{2} \cos\alpha.
\end{equation}
The total kinetic energy includes the kinetic energy due to rotation and kinetic energy due to translational motion. Rotational K.E due to rotary arm is $\dfrac{J_1}{2}\dot{\theta}^2$ and for pendulum is $\dfrac{J_2}{2}\dot{\alpha}^2$. Total rotational kinetic energy becomes

\begin{equation}
KE_1 = \dfrac{J_1}{2}\dot{\theta}^2 + \dfrac{J_2}{2}\dot{\alpha}^2.
\end{equation}
Translational kinetic energy due to rotary arm can be obtained by finding the resultant velocity of the mass ($M_1$). Resolving displacement (final-initial) of mass, $M_1$: $x_1 = \dfrac{2 L_1}{7} \cos\theta - \dfrac{2 L_1}{7}$,  $y_1 = \dfrac{2 L_1}{7} \sin\theta $, $z_1 = 0 $. Resultant velocity of mass, $M_1$: $v^2_1 = \dot{x}^2_1 + \dot{y}^2_1 + \dot{z}^2_1$. Translational K.E of rotary arm is

\begin{equation}
	\dfrac{1}{2} M_1 v^2_1 = \dfrac{2}{49} M_1 L^2_1 \dot{\theta}^2 .
\end{equation}

Translational kinetic energy due to pendulum can be obtained by finding the resultant velocity of the mass ($M_2$). Resolving translation of mass, $M_2$: $x_2 = L_1 \cos\theta + \dfrac{L_2}{2} \sin\alpha \sin\theta -L_1 $,  $y_2 = L_1 \sin\theta - \dfrac{L_2}{2} \sin\alpha \cos\theta - 0$, $z_2 = \dfrac{L_2}{2} \cos\alpha -\dfrac{L_2}{2} $ (First part of $x$ and $y$ are from rotary arm). Resultant velocity of mass, $M_2$: $v^2_2 = \dot{x}^2_2 + \dot{y}^2_2 + \dot{z}^2_2$. Translational K.E of pendulum becomes

\begin{equation}
\dfrac{1}{2} M_2 v^2_2 = 
\dfrac{M_2 L^2_2}{8} [\dot{\alpha}^2 + \sin^2\alpha \dot{\theta}^2] - \dfrac{M_2 L_1 L_2}{2} \cos\alpha \dot{\alpha}\dot{\theta} + \dfrac{M_2 L_1^2}{2}\dot{\theta}^2.
\end{equation}
Total translational K.E is obtained as

\begin{equation}
KE_2 = [\dfrac{2}{49} M_1 L^2_1 + \dfrac{M_2 L_1^2}{2}]\dot{\theta}^2 + \dfrac{M_2 L^2_2}{8} [\dot{\alpha}^2 + \sin^2\alpha \dot{\theta}^2] - \dfrac{M_2 L_1 L_2}{2} \cos\alpha \dot{\alpha}\dot{\theta}.
\end{equation}
The Lagrangian of the system is total kinetic energy - total potential energy ($L=KE_1+KE_2-PE$). Rearranging gives the relation

\begin{equation}
\begin{split}
L = \dfrac{1}{2}[J_1 + \dfrac{4}{49} M_1 L^2_1 + M_2 L^2_1 + \dfrac{1}{4}M_2 L_2^2 \sin^2\alpha] \dot{\theta}^2 + \dfrac{1}{2}[J_2 + \dfrac{1}{4}M_2 L^2_2] \dot{\alpha}^2 \\- \dfrac{1}{2}M_2 L_1 L_2 \cos\alpha \dot{\alpha} \dot{\theta} - \dfrac{1}{2}M_2 g L_2 \cos\alpha.
\end{split}
\end{equation}
Substituting in the Euler Lagrange equation, for the coordinate $\theta$ and $\alpha$ we get

\begin{equation}\label{theta}
\begin{split}
[J_1 + \dfrac{4}{49} M_1 L^2_1 + M_2 L^2_1 + \dfrac{1}{4}M_2 L_2^2 \sin^2\alpha]\ddot{\theta} - [\dfrac{1}{2} M_2 L_1 L_2 \cos\alpha] \ddot{\alpha} \\+ \dfrac{1}{4} M_2 L^2_2 \sin2\alpha \dot{\alpha} \dot{\theta}  + \dfrac{1}{2} M_2 L_1 L_2 \sin\alpha \dot{\alpha}^2 + B_1 \dot{\theta} = \tau
\end{split}
\end{equation}

\begin{equation}\label{alpha}
\begin{split}
[J_2 + \dfrac{1}{4} M_2 L^2_2] \ddot{\alpha} - [\dfrac{1}{2}M_2 L_1 L_2 \cos\alpha] \ddot{\theta} - \dfrac{1}{8}M_2 L^2_2 \sin2\alpha \dot{\theta}^2 \\ +  B_2 \dot{\alpha} - \dfrac{1}{2}M_2 g L_2 \sin\alpha = 0.
\end{split}
\end{equation}
All the parameters values associated with the rotary inverted pendulum are available at \cite{ripdatas}.
 
\section{Small angle model}
Approximating the trigonometric relations by Taylor series\cite{whittaker_taylor_cauchy}, we can reduce the non-linear terms in the Eq.\eqref{theta} and Eq.\eqref{alpha} up to the second degree. 

\begin{equation}\label{lineartheta}
\begin{split}
[J_1 + \dfrac{4}{49} M_1 L^2_1 + M_2 L^2_1]\ddot{\theta} - \dfrac{1}{2}M_2 L_1 L_2 \ddot{\alpha} + B_1 \dot{\theta} = \tau - \dfrac{1}{4}M_2 L^2_2 \alpha^2 \ddot{\theta} \\- \dfrac{1}{4}M_2 L_1 L_2 \alpha^2 \ddot{\alpha} - \dfrac{1}{2} M_2 L^2_2 \alpha \dot{\alpha} \dot{\theta} - \dfrac{1}{2}M_2 L_1 L_2 \alpha \dot{\alpha}^2
\end{split}
\end{equation}

\begin{equation}\label{linearalpha_V}
\begin{split}
[J_2 +\dfrac{1}{4}M_2L^2_2]\ddot{\alpha} - \dfrac{1}{2}M_2 L_1 L_2 \ddot{\theta}+ B_2\dot{\alpha} - \dfrac{1}{2}M_2 g L_2 \alpha = - \dfrac{1}{4}M_2 L_1 L_2 \alpha^2 \ddot{\theta} \\+ \dfrac{1}{4} M_2 L^2_2 \alpha \dot{\theta}^2
\end{split}.
\end{equation}
The torque generated by the servo motor is given by \cite{ripdatas}

\begin{equation} \label{tau}
\tau = \dfrac{\eta_g K_g \eta_m K_t (V_m - K_g K_m \dot{\theta})}{R_m}
\end{equation}

Let $\dfrac{\eta_g K_g \eta_m K_t}{R_m} = u_1$ and $\dfrac{\eta_g K_g \eta_m K_t K_g K_m}{R_m} = u_2$ then
\begin{equation}
\tau = u_1 V_m - u_2 \dot{\theta} 
\end{equation}
where $\eta_g$ is the gearbox efficiency, and $\eta_m$ is the motor efficiency with $K_g$ as high gear total gear ratio. $K_t$, $K_m$ are 
motor current torque constant and motor back emf constant respectively. $R_m$ is the motor armature resistance, $L_m$ is the motor armature inductance, and $V_m$ is the motor input voltage. Substituting $\tau$ in small angle model and rearranging to matrix form

\begin{equation} \label{matrixform}
\begin{split}
\begin{bmatrix}
J_1 + \dfrac{4}{49} M_1 L^2_1 + M_2 L^2_1 & -\dfrac{1}{2} M_2 L_1 L_2 \\ -\dfrac{1}{2} M_2 L_1 L_2 & J_2 +\dfrac{1}{4}M_2L^2_2
\end{bmatrix}
\begin{bmatrix}
\ddot{\theta} \\ \ddot{\alpha}
\end{bmatrix}
+
\begin{bmatrix}
B_1 + u_2 & 0 \\
0 & B_2
\end{bmatrix}
\begin{bmatrix}
\dot{\theta} \\ \dot{\alpha}
\end{bmatrix}
+
\begin{bmatrix}
0 & 0 \\
0 & -\dfrac{1}{2}M_2 g L_2
\end{bmatrix}
\begin{bmatrix}
\theta \\ \alpha
\end{bmatrix}
\\=
\begin{bmatrix}
u_1 \\ 0
\end{bmatrix}V_m
+
\begin{bmatrix}
- \dfrac{1}{4}M_2 L^2_2 \alpha^2 \ddot{\theta} - \dfrac{1}{4}M_2 L_1 L_2 \alpha^2 \ddot{\alpha} - \dfrac{1}{2} M_2 L^2_2 \alpha \dot{\alpha} \dot{\theta} - \dfrac{1}{2}M_2 L_1 L_2 \alpha \dot{\alpha}^2 \\
- \dfrac{1}{4}M_2 L_1 L_2 \alpha^2 \ddot{\theta} + \dfrac{1}{4} M_2 L^2_2 \alpha \dot{\theta}^2
\end{bmatrix}
\end{split}
\end{equation}
which is of the form $A \dot{X_2} + B X_2 + C X_1 = U \,V_m + N $ where $X_1 = \begin{bmatrix} \theta & \alpha \end{bmatrix}^T$, $X_2 = \dot{X_1}$, and $N$ the non-linearities.

\subsection{State Space Model}

The corresponding state model is

\begin{equation}
\begin{bmatrix}
\dot{X_1} \\ A \dot{X_2}
\end{bmatrix}
=
\begin{bmatrix}
0 & I_{2x2} \\ -C & -B
\end{bmatrix}
\begin{bmatrix}
X_1 \\ X_2
\end{bmatrix}
+
\begin{bmatrix}
0 \\ U
\end{bmatrix}V_m
+
\begin{bmatrix}
0 \\ N
\end{bmatrix}
\end{equation}
Taking $-A^{-1}C = \widetilde{C}$, $-A^{-1}B = \widetilde{B}$, $A^{-1}U = \widetilde{U}$ and $A^{-1}N = \widetilde{N}$ we obtain

\begin{equation}
\begin{bmatrix}
\dot{X_1} \\ \dot{X_2}
\end{bmatrix}
=
\begin{bmatrix}
0 & I_{2x2} \\ \widetilde{C} & \widetilde{B}
\end{bmatrix}
\begin{bmatrix}
X_1 \\ X_2
\end{bmatrix}
+
\begin{bmatrix}
0 \\ \widetilde{U}
\end{bmatrix}V_m
+
\begin{bmatrix}
0 \\ \widetilde{N}
\end{bmatrix}
\end{equation}
This takes the form

\begin{equation}
\dot{X} = A_1 X + U_1 V_m + N_1
\end{equation}
$rank(ctrb(A_1,U_1)) = 4$, hence its controllable.

\section{Controllers Design}

Let $Z(s) = R(s) - X_1(s)$ be the error signal, where $R(s) = \begin{bmatrix} \theta_d(s) & \alpha_d(s) \end{bmatrix}^T $ is the desired output. From remark~\ref{remark2}, as the system dynamics are of second-order, we require an integral, one proportional, and one derivative controller each for stabilization and control. The control input required is $V_m = K_1 Z + K_2 \dot{Z} + K_3 \int Z$. The eigenvalues of the linear part of the system can be placed towards the left half of the s-plane with the help of these gain matrices, $K_1$, $K_2$ and $K_3$, making the system stable. But the incorporation of the integral states $Z_0 = \begin{bmatrix} \int \theta - \theta_d & \int \alpha - \alpha_d \end{bmatrix}^T$ makes the system uncontrollable. Rank deficiency happens due to the integral state corresponding to $\alpha$. So from remark~\ref{remark3}, such integral states should be avoided; hence the control input takes only the integral state corresponding to $\int \theta$. \\

Controller gain values are obtained from dominant pole analysis\cite{dominantpole} along with the parameters of a second-order system. We obtain the damping ratio ($\zeta$) and undamped natural frequency ($\omega_n$) of a second-order system from the peak overshoot and settling time. So we get the real part of the dominant pole at $-\zeta \omega_n$. We take all other poles to be ten times away from the dominant pole, thereby ensuring negligible changes in the system's response.

\section{Non Linear Analysis}

Regrouping the states together in \eqref{matrixform}, we have $\dfrac{1}{4}M_2 L^2_2 \alpha^2$  and $\dfrac{1}{4}M_2 L_1 L_2 \alpha^2$ in matrix $A$ from the non-linear part. As the action of controller starts close to $\alpha = 0$, compared to the other terms they are negligible. Ignoring these and taking the inverse of $A$ and we obtain

\begin{equation} \label{dynamicsform}
\begin{split}
\begin{bmatrix}
\ddot{\theta} \\ \ddot{\alpha}
\end{bmatrix}
=
\begin{bmatrix}
v_1 \\ v_2
\end{bmatrix}V_m
-
\begin{bmatrix}
b_{11} & b_{12} \\
b_{21} & b_{22}
\end{bmatrix}
\begin{bmatrix}
\dot{\theta} \\ \dot{\alpha}
\end{bmatrix}
-
\begin{bmatrix}
0 & c_1 \\
0 & c_2
\end{bmatrix}
\begin{bmatrix}
\theta \\ \alpha
\end{bmatrix}
+
\begin{bmatrix}
a_1 \alpha \dot{\alpha} \dot{\theta} + a_2 \alpha \dot{\alpha}^2 + a_3 \alpha \dot{\theta}^2 \\ a_4 \alpha \dot{\alpha} \dot{\theta} + a_5 \alpha \dot{\alpha}^2 + a_6 \alpha \dot{\theta}^2 
\end{bmatrix}
\end{split}
\end{equation}
where $A^{-1} = \begin{bmatrix}
289.1545 & 278.1123 \\
278.1123 & 475.5730
\end{bmatrix}, 
\begin{bmatrix}
b_{11} & b_{12} \\ b_{21} & b_{22}
\end{bmatrix} = 
\begin{bmatrix}
20.6543 & 0.6675 \\
19.8655 & 1.1414
\end{bmatrix}
$,
\\

$\begin{bmatrix}
c_1 \\ c_2
\end{bmatrix} = 
\begin{bmatrix}
-58.3839 \\ -99.8366
\end{bmatrix}, 
\begin{bmatrix}
v_1 \\ v_2
\end{bmatrix} = 
\begin{bmatrix}
37.1285 \\ 35.7106
\end{bmatrix}
$ and \\

$a_1 = -2.0852$, $a_2 = -1.3366$, $a_3 = 1.0028$, $a_4 = -2.0056$, $a_5 = -1.2855$, $a_6 = 1.7148$

\subsection{Non Linear Dynamics}

Let $x_1 = \theta$, $x_2 =\alpha$, $x_3 = \dot{\theta}$, $x_4 = \dot{\alpha}$, and adding the integral state $x_0 = \int \theta$, the state equation becomes

\begin{equation}\label{stateeq}
\begin{split}
\dot{x_0} &= x_1 \\
\dot{x_1} &= x_3 \\
\dot{x_2} &= x_4 \\
\dot{x_3} &= v_1 V_m - b_{11}x_3 - b_{12}x_4 - c_1 x_2 + a_1 x_2 x_3 x_4 + a_2 x_2 x_4^2 + a_3 x_2 x_3^2 \\
\dot{x_4} &= v_2 V_m - b_{21}x_3 - b_{22}x_4 - c_2 x_2 + a_4 x_2 x_3 x_4 + a_5 x_2 x_4^2 + a_6 x_2 x_3^2.\
\end{split}
\end{equation}
By keeping the reference to zero, the error state becomes the same as the system states. Incorporating the state feedback $-V_m = k_0 x_0 + k_1 x_1 + k_2 x_2 + k_3 x_3 + k_4 x_4$, the dynamics takes the form $\dot{X} = A_d X + N_d$. $A_d$ is the refined state matrix, which is composed of only the linear terms and $N_d$ the non-linearities associated with the system.

\begin{equation}
A_d = \begin{bmatrix}
0 & 1 & 0 & 0 & 0\\
0 & 0 & 0 & 1 & 0 \\
0 & 0 & 0 & 0 & 1 \\
-v_1k_0 & -v_1 k_1 & -(v_1 k_2 +c_1) & -(v_1 k_3 + b_{11}) & -(v_1 k_4 + b_{12}) \\
-v_2 k_0 & -v_2 k_1 & -(v_2 k_2 +c_2) & -(v_2 k_3 + b_{21}) & -(v_2 k_4 + b_{22})
\end{bmatrix}
\end{equation}
negative eigenvalues can be obtained by choosing proper values of gain matrix, hence making the equilibrium points stable.

\subsection{Boundedness}

Without of loss of generality we take the controller input as error signal with zero reference, hence Z(t) = X(t). 
 
\begin{equation} \label{statesolution}
Z(t)=e^{A_d(t)}\,Z(0)+\int\limits_{0}^{t}e^{A_d(t-\tau)}\,N_d(z,\tau)\,d\tau.
\end{equation}
Taking the Euclidean norm and applying triangular inequality

\begin{equation}
\|Z(t)\| \leq \|e^{A_d t} Z(0)\| + \| \int\limits_{0}^{t}e^{A_d(t-\tau)}\,N_d(z,\tau)\,d\tau\|.
\end{equation}
We can diagonalize $A_d$ by a similarity transformation $M \Sigma M^{-1}$, where $M$ is the model matrix whose columns are eigenvectors.

\begin{equation}
e^{A_d} = M e^{\Sigma} M.
\end{equation} 
Note $\Sigma$ has its diagonal entries as the eigenvalues ($\lambda$) such that $\lambda_1 \geq \lambda_2 \geq \lambda_3 \geq \lambda_4 \geq \lambda_5$. These eigenvalues can be made distinct by a proper selection of gain matrix. Taking the norm and substituting a general constant $ \beta = \|M\| \|M^{-1}\| $, ($\beta =1$ for orthogonal matrix, $M$)

\begin{equation}
\| e^{A_d} Z(0) \| \leq  \beta \| e^{\max(\lambda)} Z(0) \| = \beta \| e^{\lambda_1 t} Z(0)\|.
\end{equation}
As all the eigenvalues are negative, $e^{\lambda}$ can at most attain 1, hence

\begin{equation}
\| e^{A_d} Z(0) \| \leq \beta \| Z(0) \|
\end{equation}
substituting back and rewriting 
\begin{equation}
\|Z(t)\| \leq \beta \| Z(0) \| + \beta \int\limits_{0}^{t} \| e^{\lambda_1(t-\tau)}\| \, \|N_d(z,\tau) \| \,d\tau.
\end{equation}
From Eq.\eqref{stateeq}, norm of the non-linearities can be seen as \\
$\| N_d(x) \| = \sqrt{\left( a_1 z_2 z_3 z_4 + a_2 z_2 z_4^2 + a_3 z_2 z_3^2 \right)^2 + \left( a_4 z_2 z_3 z_4 + a_5 z_2 z_4^2 + a_6 z_2 z_3^2 \right)^2 } $. As $\|Z \| = \sqrt{z_0^2 + z_1^2 + z_2^2 + z_3^2 + z_4^2} $ and $\|z_i \| \leq \|Z \|$, we have

\begin{equation}
\| N_d(z,t) \| \leq \kappa\|Z(t) \|^3 
\end{equation} 
where $\kappa = \sqrt{(a_1+a_2+a_3)^2 + (a_4+a_5+a_6)^2} $. Let $\|Z(t)\|$ is bounded by constant $\gamma$, then 

\begin{equation}
\begin{split}
\int\limits_{0}^{t} \| e^{A_d(t-\tau)} \|  \| \, N_d(z,\tau) \|  \,d\tau &\leq \beta \, \kappa\, \gamma^3 \| \int\limits_{0}^{t} e^{\lambda_1(t-\tau)} \, d\tau \| \\ &\leq \beta \, \kappa\, \gamma^3 \| \dfrac{e^{\lambda_1} - 1}{\lambda_1} \| \leq \dfrac{\beta \, \kappa\, \gamma^3}{|\lambda_1|}
\end{split}
\end{equation}
combining both terms we have

\begin{equation}
\|Z(t)\| \leq \beta \| Z(0) \| + \dfrac{\beta \, \kappa\, \gamma^3}{|\lambda_1|} \leq \gamma
\end{equation}
for small initial conditions $\gamma^3$ will be less than $\gamma$, and our states never gets out of the bound.

\begin{figure}[!h]
\begin{center}
\includegraphics[scale=.15]{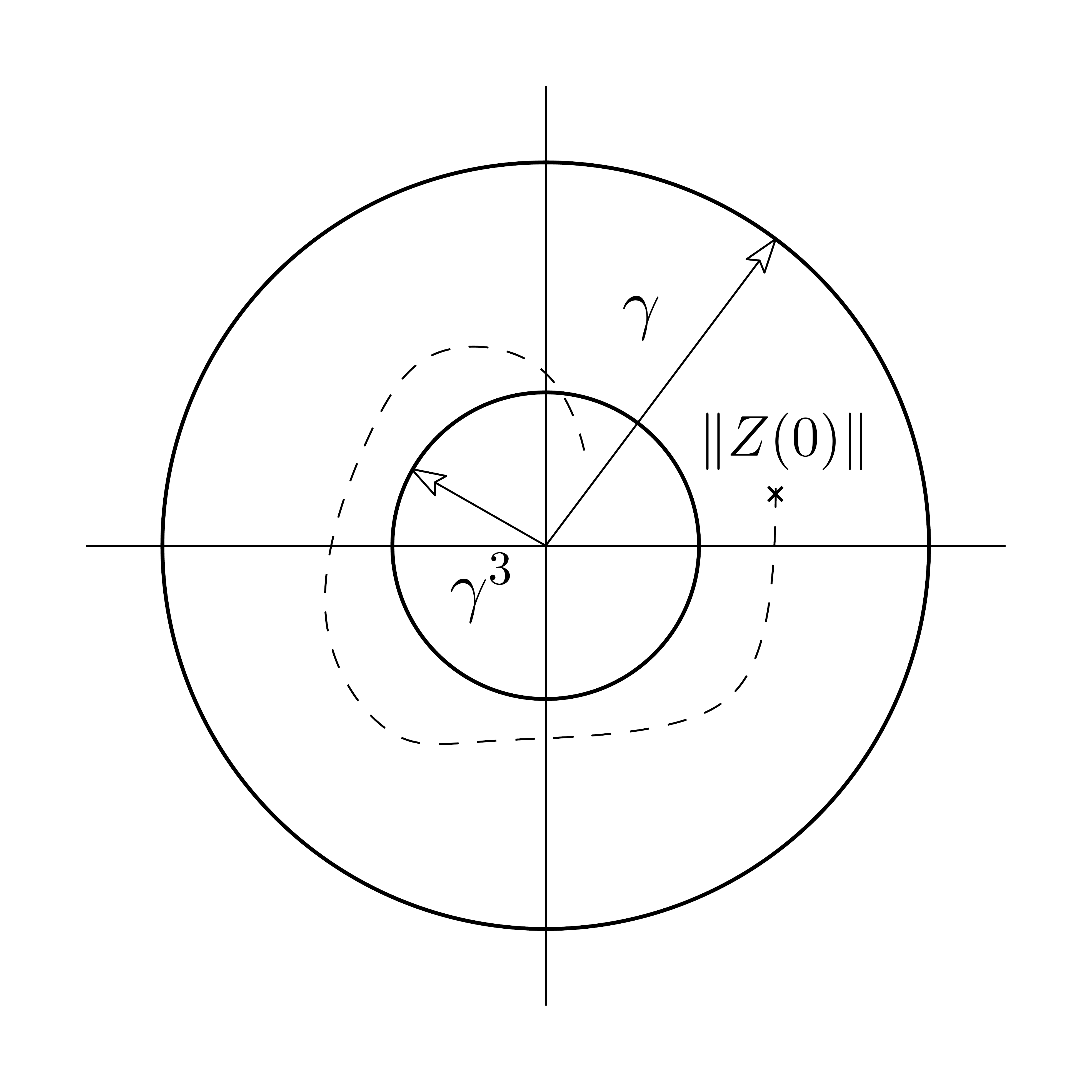}
\caption{Illustration of bounds on states and non-linearity}
\label{rip_bounds}
\end{center}
\end{figure}

\vspace{-1cm}
\subsection{Convergence}

To show that the system is stabilized with the controller, convergence is also required along with boundedness. This can be seen by taking the state at two instants of time, and showing its norm of difference decreases over time. From Eq.\eqref{statesolution}

\begin{equation} 
Z(t_1)=e^{A_d(t_1)}\,Z(0)+\int\limits_{0}^{t_1}e^{A_d(t_1-\tau)}\,N_d(z,\tau)\,d\tau
\end{equation}

\begin{equation} 
Z(t_2)=e^{A_d(t_2)}\,Z(0)+\int\limits_{0}^{t_2}e^{A_d(t_2-\tau)}\,N_d(z,\tau)\,d\tau
\end{equation}
where $t_1 = t_2 + t_0$, and $t_0$, $t_1$, $t_2$ $\in \mathbb{N}$

\begin{equation} 
\begin{split}
Z(t_1)=e^{A_d(t_2)}e^{A_d(t_0)}\,Z(0) + \int\limits_{0}^{t_2}e^{A_d(t_2-\tau)}e^{A_d(t_0)}\,N_d(z,\tau)\,d\tau \\+ \int\limits_{t_2}^{t_2+t_0}e^{A_d(t_2-\tau)}e^{A_d(t_0)}\,N_d(z,\tau)\,d\tau
\end{split}
\end{equation}
taking the norm of the difference and substituting the upper bound for non linear term

\begin{equation} \label{converge}
\begin{split}
\|Z(t_1) - Z(t_2)\| \leq \|e^{A_d(t_2)} \left(e^{A_d(t_0)} - I\right)\| \, \|Z(0)\| + \int\limits_{0}^{t_2} \|e^{A_d(t_2-\tau)} \left(e^{A_d(t_0)} - I\right)\| \,\gamma^3 d\tau \\+ \int\limits_{t_2}^{t_2+t_0} \|e^{A_d(t_2-\tau)}e^{A_d(t_0)}\| \, \gamma^3 d\tau
\end{split}
\end{equation}
note that all the terms except terms with $t_2$ are constants, and $e^{A_d}$ is bounded by exponential of maximum of eigenvalue. Evaluating the integral

\begin{equation} \label{integralbound}
\begin{split}
\| \int\limits_{0}^{t_2} e^{A_d(t_2-\tau)} d\tau \| \leq \| \int\limits_{0}^{t_2} e^{\lambda_{max}(t_2-\tau)} d\tau \| \\ \leq \| \dfrac{e^{\lambda_{max}t_2} -1}{\lambda_{max}} \|
\end{split}
\end{equation}
as all the eigenvalues are negative, its a finite value, hence

\begin{equation}
\| Z(t_1) - Z(t_2)\| \leq \varepsilon
\end{equation}
Note that the integral term in Eq.\eqref{statesolution} is bounded by $\dfrac{1}{\lambda_{max}}$. $Z(t)$ always reduce if the initial states are small enough. This in turns reduce the non-linear term and we have in Eq.\eqref{converge}, as $t_1$, $t_2$ $\longrightarrow \infty$, RHS $\longrightarrow 0 $. So we get a Cauchy sequence\cite{whittaker_taylor_cauchy} and the states converges over time.

\section{Experimental Setup}

The Quanser rotary inverted pendulum is clamped  at the corner of the laboratory desk, such that the pendulum arm is free to move. The rotary arm stabilizes the pendulum, which is actuated by faulhber coreless DC motor (2338S006 series) which has high efficiency, and low inductance for faster response. The nominal voltage rating is $\SI{6} {\volt}$. It can withstand $\pm \SI{15} {\volt}$, $\SI{1} {\ampere}$ continuous current and $\SI{3} {\ampere}$ peak current. The angles $\alpha$ and $\theta$ are measured using two encoders which are present at the end of rotary arm and at the fixed base respectively. Rotary arm encoder resolution and pendulum arm is set to $ 2\pi / (4*1024)$  $rad/count$. The measured signals are fed via the data acquisition board (DAB) to the computer. The DAB drives the actuator through the power amplifier. Here the external gear configuration is set to 'HIGH' and amplifier gain is set to 1. 

\begin{figure}[!h]	
	\advance \leftskip-2cm
	\begin{center}
	\includegraphics[scale=.4]{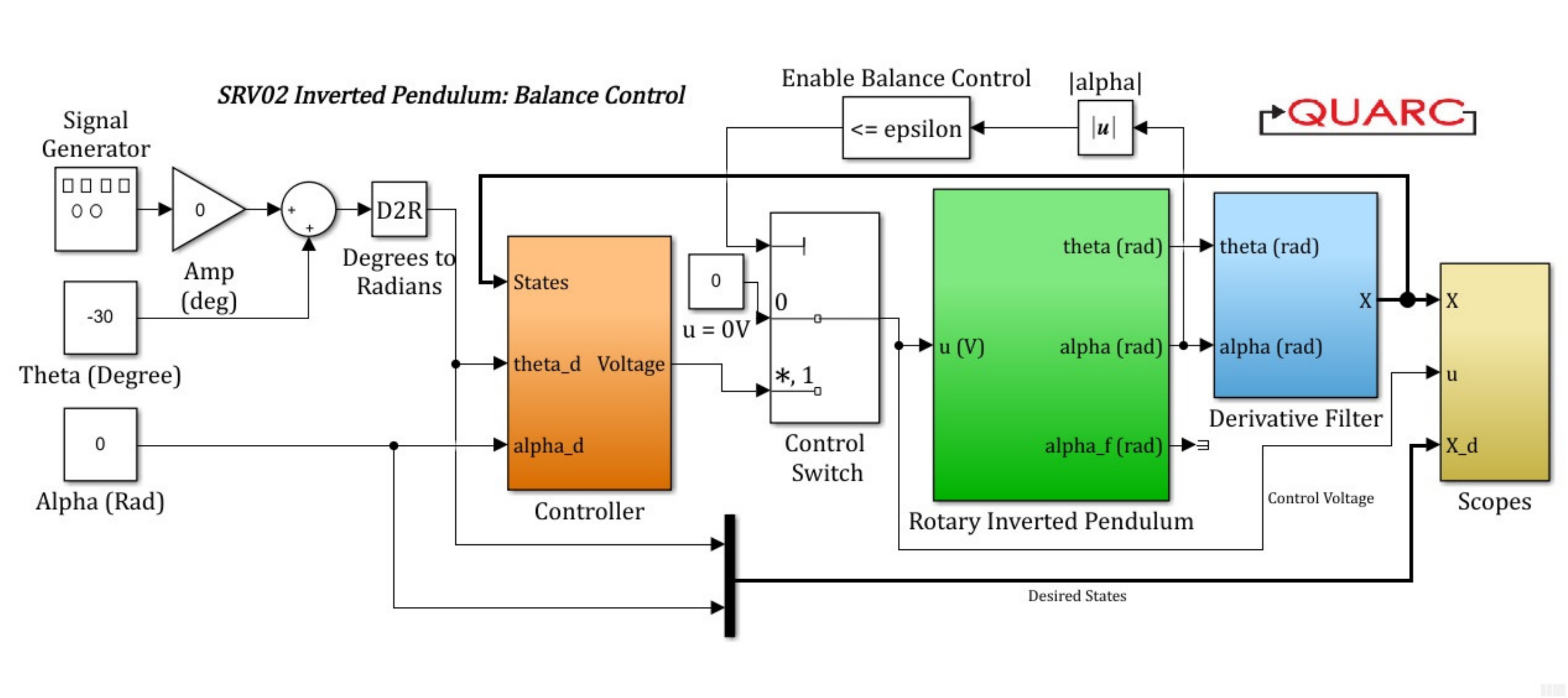}
	\caption{PID and PD controllers with inverted pendulum system.}
	\label{2dof_controller_block}
	\end{center}
\end{figure}

Interfacing between the hardware and software (MATLAB) is done using the QUARC hardware block, and QUARC library. Due to mechanical constrain  the rotor angle is limited to $\pm \SI{45} {\degree}$. Derivative states are obtained from the outputs using the derivative block, accompanied by a low pass filter with a cutoff frequency of $20 \pi$ $Hz$. Integral windup occurs due to the use of an integral controller. Hence a back-calculation anti integral windup\cite{antiwindup} with an integral reset time of $\SI{1}{\second}$ is used along with the integral controller to eliminate it.

\begin{figure*}[!h]
\advance\leftskip-2cm
\centering
\includegraphics[scale=.4]{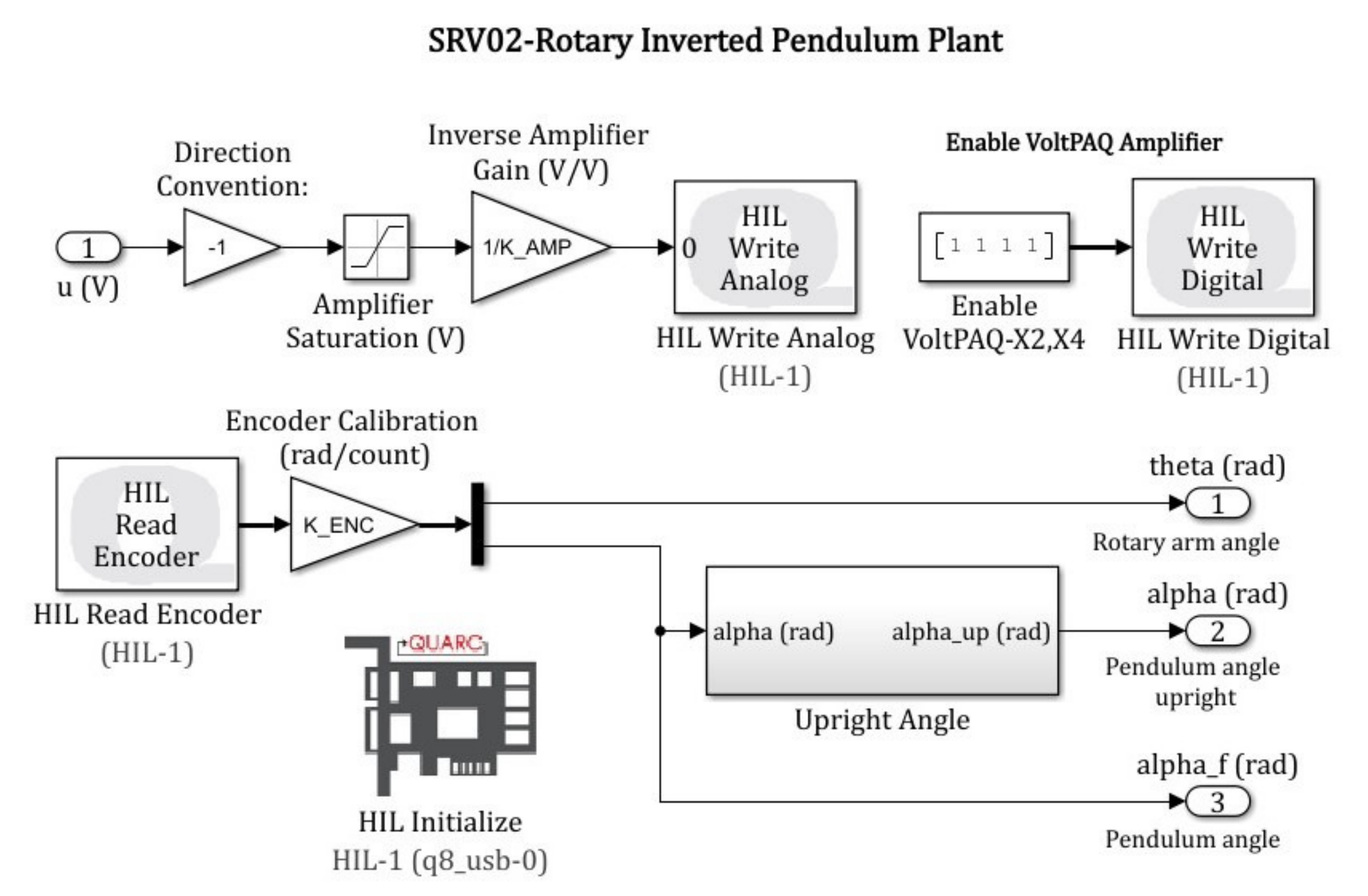}
\caption{Hardware data management block.}
\label{2dof_hardware}
\end{figure*}

\subsection{Pole Placement}

For a second order characteristics of $2\%$ overshoot and $.7797$ damping ratio, the poles are placed at $-2 \pm 1.606i$, $-10$, $-12$, and $-15$. Which gives the gain values as

\begin{equation}
K = \begin{bmatrix}
-7.302 & -6.348 & 27.681 & -3.166 & 3.829
\end{bmatrix}
\end{equation}
which is basically the PD controller for $\alpha$ and PID controller for $\theta$.

\subsection{Results}
 
\begin{figure}[!h]
\advance\leftskip-1cm
\centering
\includegraphics[scale=.3]{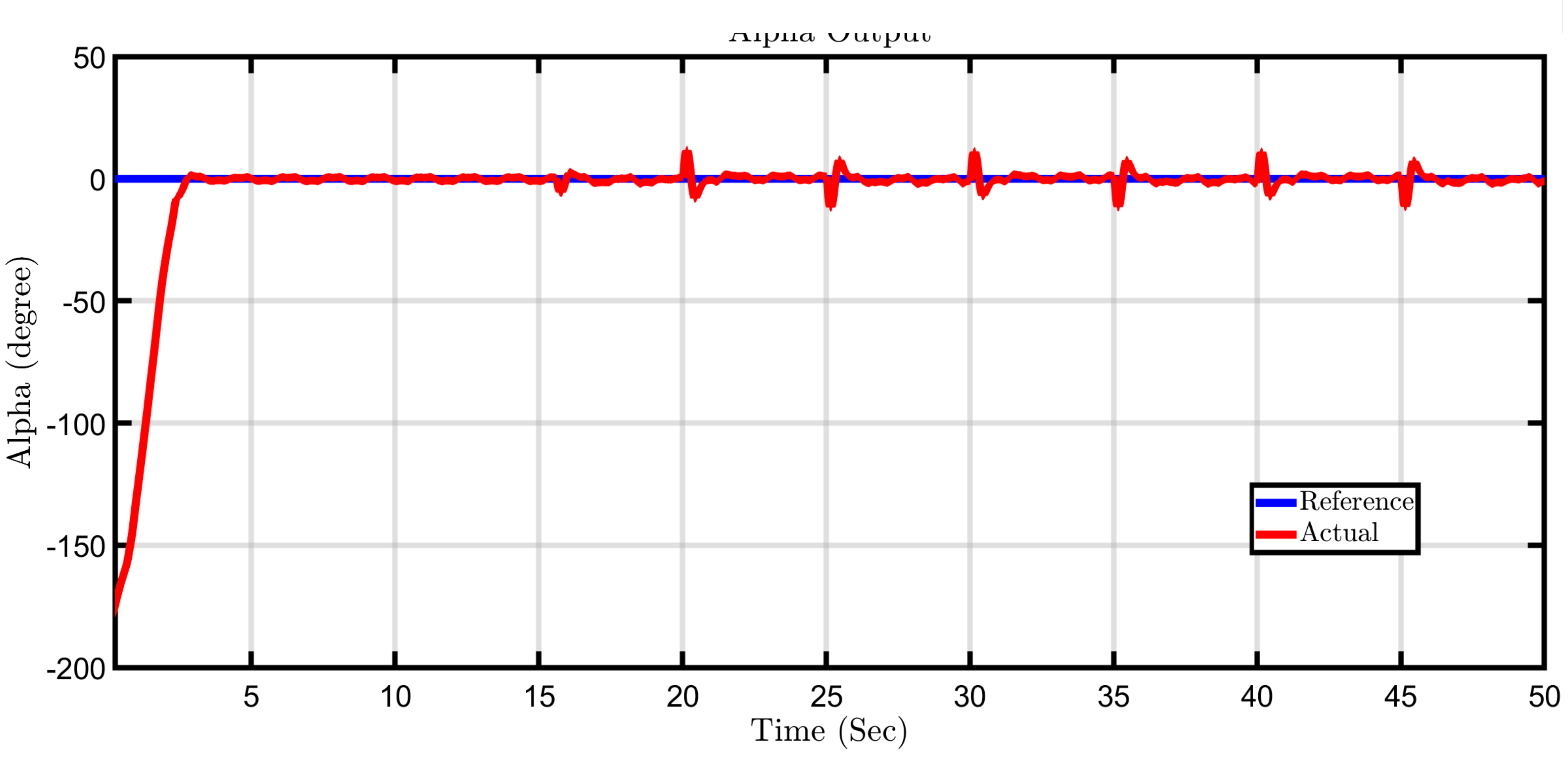}
\caption{Pendulum position w.r.t vertical}
\label{rip_result21}
\end{figure}

\begin{figure}[!h]
\advance\leftskip-1cm
\centering
\includegraphics[scale=.3]{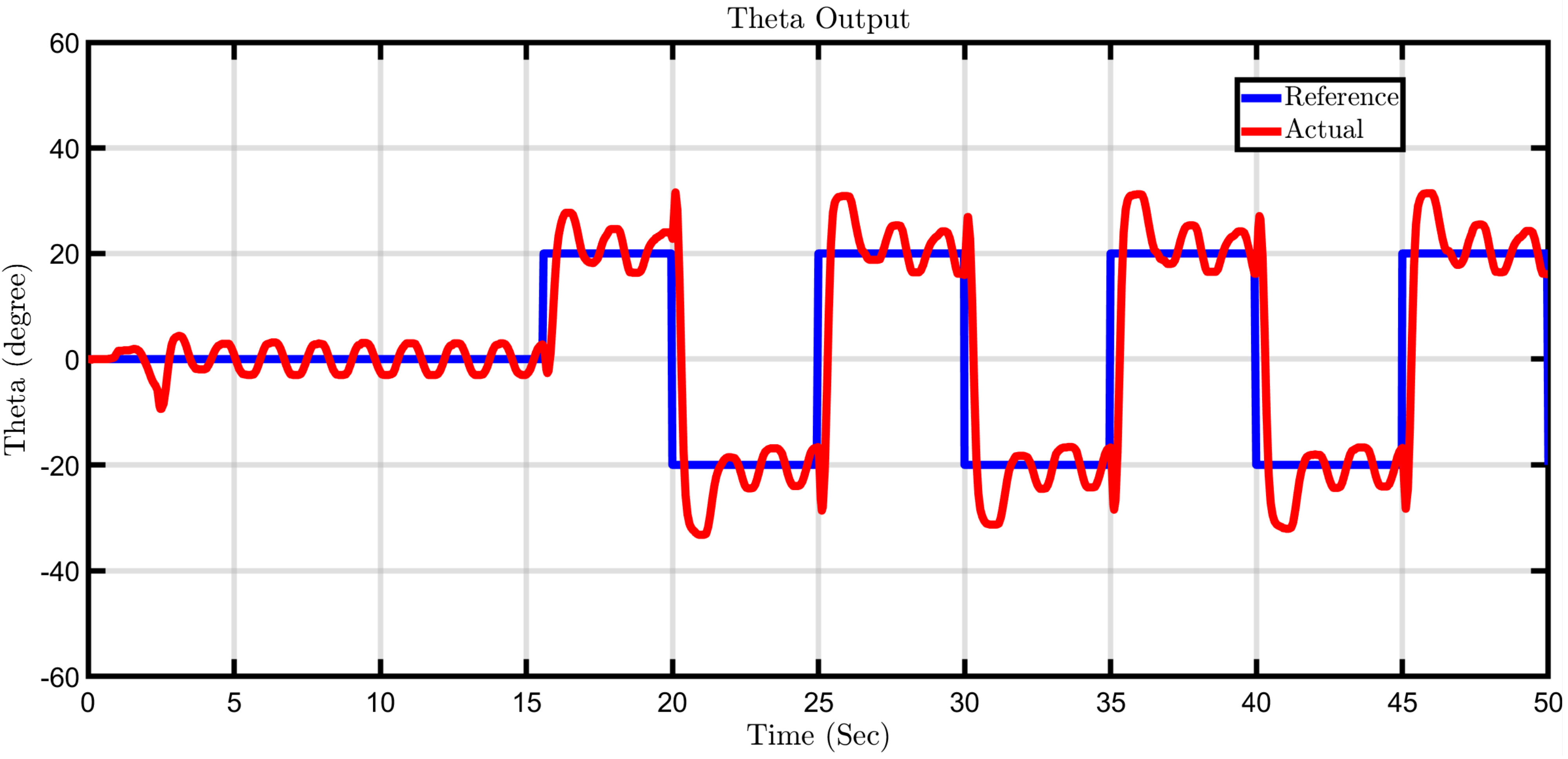}
\caption{Rotary arm position}
\label{rip_result22}
\end{figure}

\begin{figure}[!h]
\advance\leftskip-1cm
\centering
\includegraphics[scale=.3]{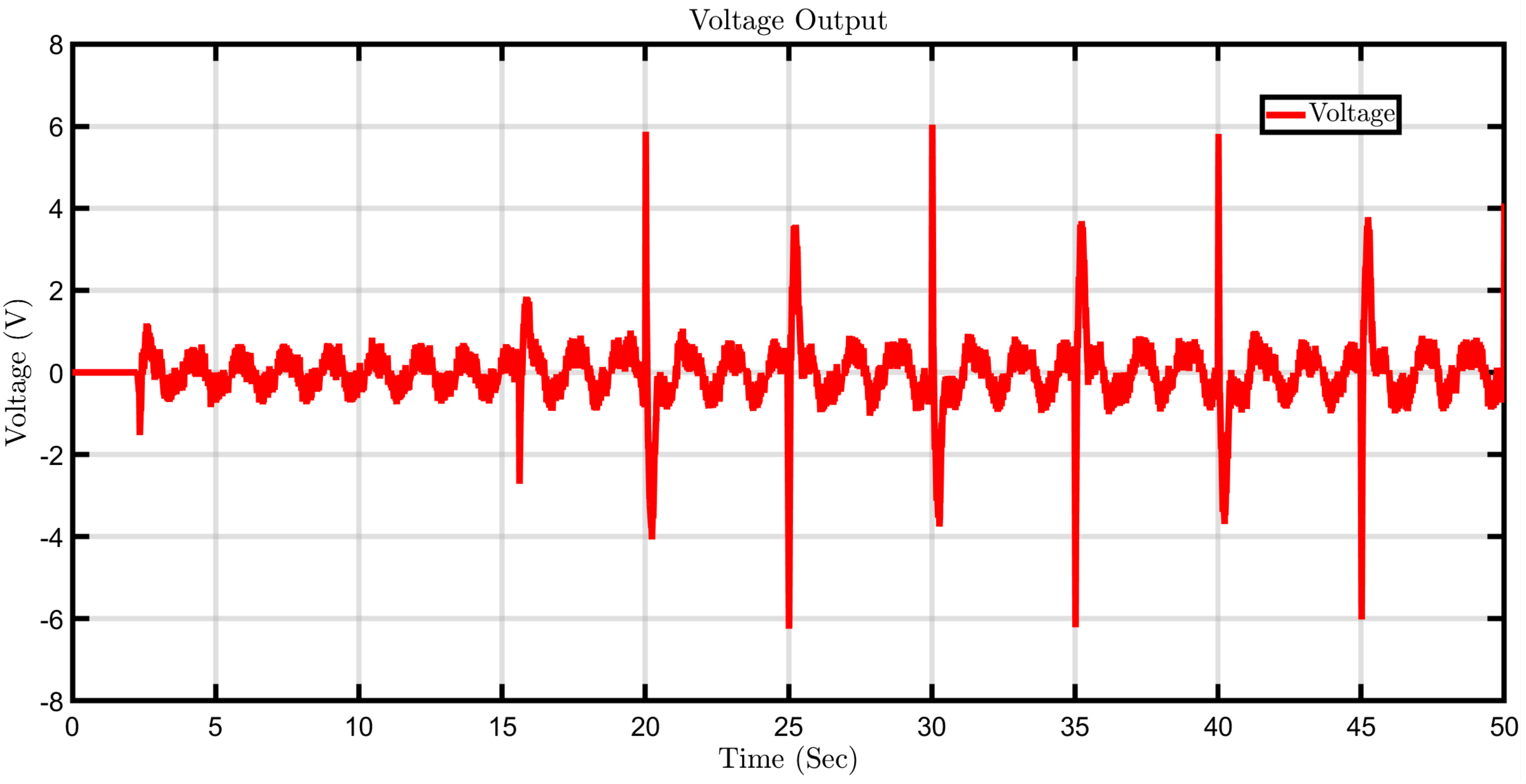}
\caption{Rotary arm motor voltage}
\label{rip_result23}
\end{figure}

Initially the pendulum arm is brought to upright position and the control action starts when the pendulum angle $\alpha$ is $\leq$ $\pm \SI{20} {\degree}$. For the initial period of $\SI{15}{\second}$ the pendulum is stabilized for $\theta = \SI{0} {\degree}$. For the later half of experiment a square pulse is given as the reference $\theta$, which varies $\theta$ between $+ \SI{20} {\degree} $ to $ - \SI{20} {\degree}$. with a time period of $\SI{10}{\second}$. The experiment results for a total duration of $\SI{50}{\second}$ is presented here. Figure \ref{rip_result21} shows the reference, and actual output angle of the pendulum arm w.r.t vertical in degrees. Figure \ref{rip_result22} shows the reference, and actual output of the rotary arm angle in degrees. The actual voltage output to the rotary arm motors, is shown in figure \ref{rip_result23}.

\subsection{Inference}

Since the system is highly non-linear and coupled, a minor disturbance will cause the system to oscillate. Increasing the gain value can reduce the oscillation peaks in $\theta$, which is constrained by the maximum voltage applied to the motor. Gain values are designed such that the overshoot in the simulations is negligible. Since we use a derivative controller, a low pass filter must accompany the controller; else, it may damage the actuator. The proposed controller scheme obtains the controller parameters easily, as the key idea is how the gains are related to the system dynamics and desired characteristic coefficient.

\section{Conclusion}

New control strategies are usually experimented first on fundamental systems. Here we have extended a linear controller scheme for the application of a non-linear system and have shown its effectiveness in stabilizing the rotary inverted pendulum. The proposed linear controller is able to reject any bounded disturbance acting on the system due to the integral controller part in it.  The key idea in extending the controller application to non-linear systems is to see the non-linearity as the disturbance acting on the linear system. Any unstable operating point can be stabilized by pole placement if the system is controllable. Since the controller parameters depend only on the linear system characteristic coefficients and the desired characteristic coefficients, it's straightforward to design the gain matrix. The experimental results for stabilizing the pendulum arm and controlling the rotary arm are successfully demonstrated. 

\bibliography{reference}
\bibliographystyle{elsarticle-num}

\end{document}